%% file: main.tex
\documentclass[twocolumn, tighten, times, twocolappendix]{aastex631}
\let\tablenum\relax
\usepackage{siunitx}
\usepackage{amsmath}
\usepackage{microtype}
\graphicspath{{figures/}}

\input{defs}

\shorttitle{A revised energy formalism for common-envelope evolution}
\shortauthors{Yarza et al.}

\begin{document}

\title{A revised energy formalism for common-envelope evolution: repercussions for planetary engulfment and the formation of neutron star binaries}

\author[0000-0003-0381-1039]{Ricardo Yarza}
\altaffiliation{NASA FINESST Fellow}
\altaffiliation{Frontera Computational Science Fellow}
\affiliation{Department of Astronomy and Astrophysics, University of California, Santa Cruz, CA 95064, USA}
\affiliation{Texas Advanced Computing Center, University of Texas, Austin, TX 78759, USA}

\author[0000-0001-5256-3620]{Rosa Wallace Everson}
\altaffiliation{NSF Graduate Research Fellow}
\affiliation{Department of Astronomy and Astrophysics, University of California, Santa Cruz, CA 95064, USA}

\author[0000-0003-2558-3102]{Enrico~Ramirez-Ruiz}
\affiliation{Department of Astronomy and Astrophysics, University of California, Santa Cruz, CA 95064, USA}

\correspondingauthor{Ricardo Yarza}
\email{ryarza@ucsc.edu}

\input{abstract}

\input{introduction}
\input{derivation}
\input{comparison}
\input{methods}

\section{Discussion}
\label{sec:discussion}
\input{ns_binaries}

\input{planetary_engulfment}

\input{conclusion}

\begin{acknowledgments}
We thank Alejandro Vigna-G\'{o}mez for discussions about COMPAS, and Tenley Hutchinson-Smith for discussions about hydrodynamical simulations of common-envelope evolution. We used the lux supercomputer at UC Santa Cruz, funded by NSF MRI grant AST 1828315. We thank Brant Robertson and Josh Sonstroem for help using lux. R.Y. is grateful for support from a Doctoral Fellowship from UCMEXUS and CONACyT, a Frontera Computational Science Fellowship from the Texas Advanced Computing Center, and a NASA FINESST Fellowship (21-ASTRO21-0068). R.W.E. is supported by the NSF Graduate Research Fellowship Program (Award \#1339067), the Heising-Simons Foundation, and the Vera Rubin Presidential Chair for Diversity at UCSC. E. R-R thanks the Heising-Simons Foundation and the NSF (AST-1911206, AST-1852393, and AST-1615881) for support.
\end{acknowledgments}

\software{GNU Scientific Library 2.7 \citep{gsl}, HDF5 \citep{hdf5}, matplotlib 3.6.0 \citep{matplotlib}, MESA r22.05.1 \citep{Fuller1985,Iglesias1993,Oda1994,Saumon1995,Iglesias1996,Itoh1996,Angulo1999,Langanke2000,Timmes2000,Rogers2002,Irwin2004,Ferguson2005,Cassisi2007,Chugunov2007,Cyburt2010,Potekhin2010,Paxton2011,Paxton2013,Paxton2015,Poutanen2017,Paxton2018,Paxton2019,Blouin2020,Jermyn2021}, MESA SDK 22.6.1 \citep{Townsend2021}, numpy 1.23.3 \citep{Harris2020}, py\_mesa\_reader 0.3.0 \citep{pymesareader}, scipy 1.9.1 \citep{scipy}.}

\input{appendix}

\bibliography{bib}

\end{document}

%% file: defs.tex
\newcommand{\lp}{\left(}
\newcommand{\rp}{\right)}
\newcommand{\ls}{\left[}
\newcommand{\rs}{\right]}

\newcommand{\lb}{\left\{}
\newcommand{\rb}{\right\}}

\newcommand{\menc}{\ensuremath{M_\text{enc}}}
\newcommand{\menczero}{\ensuremath{M_{\text{enc},0}}}

\newcommand{\eorb}{\ensuremath{E_\text{orb}}}

%% file: abstract.tex
\begin{abstract}
\noindent Common-envelope evolution is a stage in binary system evolution in which a giant star engulfs a companion. The standard energy formalism is an analytical framework to estimate the amount of energy transferred from the companion's shrinking orbit into the envelope of the star that engulfed it. We show analytically that this energy transfer is larger than predicted by the standard formalism. As the orbit of the companion shrinks, the mass it encloses becomes smaller, and the companion is less bound than if the enclosed mass had remained constant. Therefore, more energy must be transferred to the envelope for the orbit to shrink further. We derive a revised energy formalism that accounts for this effect, and discuss its consequences in two contexts: the formation of neutron star binaries, and the engulfment of planets and brown dwarfs by their host stars. The companion mass required to eject the stellar envelope is smaller by up to $50\%$, leading to differences in common-envelope evolution outcomes. The energy deposition in the outer envelope of the star, which is related to the transient luminosity and duration, is up to a factor of $\approx7$ higher. Common-envelope efficiency values above unity, as defined in the literature, are thus not necessarily unphysical, and result at least partly from an incomplete description of the energy deposition. The revised energy formalism presented here can improve our understanding of stellar merger and common-envelope observations and simulations.
\end{abstract}

%% file: introduction.tex
\section{Introduction}
Common-envelope evolution \citep[hereafter CEE;][]{Paczynski1976} occurs when a post-main-sequence star fills its Roche lobe and engulfs its companion. This process is thought to shape the fates of a significant fraction of binary stars, leading to the formation of systems such as compact object and white dwarf binaries, the conjectured Thorne-Zytkow objects, and millisecond pulsars \citep{Ivanova2013a}.

Similarly, many planets and brown dwarfs will undergo common-envelope evolution with their host stars \citep{Nordhaus2013}---a process more often referred to as planetary engulfment. This process has been studied as a potential explanation for several observations in stellar and planetary astrophysics, such as substellar companions around stellar remnants \citep{Livio1984,Nelemans1998,Staff2016,Kramer2020,Bear2021,Yarza2022}, and populations of giant stars with abnormal rotation rates \citep[e.g.,][]{Sandquist1998,Siess1999,Sandquist2002,AguileraGomez2016,AguileraGomez2016a, SoaresFurtado2021} and surface abundances of the $^7$Li lithium isotope \citep[e.g.,][]{Sandquist1998,Siess1999,Sandquist2002,AguileraGomez2016,AguileraGomez2016a,SoaresFurtado2021}.

The complexity of CEE---characterized by the wide range of physical processes, and spatial and temporal scales involved---, combined with the need for computationally inexpensive prescriptions for population synthesis, has motivated the development of simplified analytical frameworks to understand it. One of such frameworks is the ``standard energy formalism'' \citep{vandenHeuvel1976,Webbink1984}, based on the transfer of energy from the orbit of the companion to the envelope of the star that engulfed it. This formalism is used to estimate whether envelope ejection is possible, by comparing the change in orbital energy to the binding energy of the envelope.

The standard energy formalism suggests that, as a result of energy conservation, the change in orbital energy is equal to the work done on the companion, and therefore equal in magnitude to the energy deposited in the envelope. This deposited energy can be written as
\begin{equation}
    \label{eq:standard_formalism}
    E_\text{dep,st} = - \Delta\eorb,
\end{equation}
where ``st'' stands for ``standard,'' and $\Delta\eorb$ is the change in orbital energy.

This Letter shows that equation \ref{eq:standard_formalism} does not hold because the mass enclosed by the orbit of the companion is a function of time. In \ref{sec:derivation}, we derive a revised energy formalism (in the limit where the companion mass is much smaller than the mass enclosed by its orbit) and show that conservation of energy requires that the energy deposited into the envelope is larger in magnitude than the change in orbital energy. In \ref{sec:comparison}, we compare this revised formalism to the standard one. Sections \ref{sec:discussion:ns} and \ref{sec:discussion:planet} discuss the implications for CEE involving compact and substellar companions, respectively. We summarize our results in \ref{sec:conclusion}. The Appendix contains fitting formul\ae{} for some of our results and a reference to the reproducibility repository for this Letter.

%% file: derivation.tex
\section{A revised energy formalism}\label{sec:derivation}
We consider an extended star (``primary'') engulfing a companion of mass $M_2$. We assume that the mass of the companion is much smaller than the mass enclosed by its orbit, $\menc$. In this limit, the equation of motion for the companion is
\begin{equation}
    \label{eq:eom}
    M_2 \frac{d\mathbf{v}}{dt} = M_2 \mathbf{g} + \mathbf{F_d},
\end{equation}
where $\mathbf{v}$ is its velocity, $t$ is time, $\mathbf{g}$ is the gravitational field and $\mathbf{F_d}$ is the drag force. The gravitational field is
\begin{equation}
    \mathbf{g}=-\frac{G M_\text{enc}}{r^2}\mathbf{\hat{r}},
\end{equation}
where $r$ is the radial coordinate of the companion with respect to the center of the primary, $\mathbf{\hat{r}}$ is the unit vector in the radial direction, and $G$ is the gravitational constant. We will write an equation for the rate of change of the orbital energy of the companion, which equals the sum of its potential and kinetic energies. Taking the dot product of the equation of motion \ref{eq:eom} and the velocity of the companion yields the equation for the rate of change of its kinetic energy,
\begin{equation}
    \label{eq:ekin}
    \frac{dE_\text{kin}}{dt}= M_2 \mathbf{v}\cdot\mathbf{g}+\mathbf{F_d}\cdot \mathbf{v}=-\frac{G M_\text{enc}M_2\dot{r}}{r^2}+\mathbf{F_d}\cdot \mathbf{v},
\end{equation}
which is the work-energy theorem for the companion. The first term on the right-hand side (RHS) is the work per unit time done by the gravitational force $M_2\mathbf{g}$. Since the gravitational field points radially inwards, only the radial component of the velocity $\dot{r}=\mathbf{v}\cdot\mathbf{\hat{r}}$ contributes to this term. The second term on the RHS is the work per unit time done by the drag forces responsible for the orbital decay of the companion.

The potential energy is
\begin{equation}
    E_\text{pot}=-\frac{G M_2 M_\text{enc}}{r},
\end{equation}
and its rate of change is
\begin{equation}
    \label{eq:epot}
    \frac{dE_\text{pot}}{dt}=\frac{G M_2 M_\text{enc} \dot{r}}{r^2}-\frac{G M_2 \dot{r}}{r}\frac{dM_\text{enc}}{dr}.
\end{equation}
The second term on the RHS is the contribution from the changing enclosed mass. Adding equations \ref{eq:ekin} and \ref{eq:epot}, we obtain
\begin{equation}
    \label{eq:deorbdt}
    \frac{dE_\text{orb}}{dt}=\frac{dE_\text{pot}}{dt}+\frac{dE_\text{kin}}{dt}=-\frac{G M_2 \dot{r}}{r}\frac{dM_\text{enc}}{dr}+\mathbf{F_d}\cdot \mathbf{v}.
\end{equation}
If the enclosed mass is constant and there are no drag forces, the RHS is zero, so the orbital energy of the companion is conserved. Conservation of orbital energy is expected in that limit because the equations then represent the Keplerian motion of a test particle. If the companion experiences drag forces, but the enclosed mass is constant, the orbital energy changes according to the work-energy theorem, with the drag force adding or subtracting energy from the companion depending on its alignment with the velocity. However, if the enclosed mass changes with time, there is an additional term whose sign depends on the rate of change of the enclosed mass.

Replacing $t$ with the integration variable $t'$ and integrating from $t'=0$ to $t$ yields
\begin{equation}
    \label{eq:int}
    \int_0^t \frac{dE_\text{orb}}{dt'}dt'=-G M_2 \int_0^t \frac{\dot{r}}{r}\frac{dM_\text{enc}}{dr}dt'+W_d,
\end{equation}
where we have defined the work done by the drag forces
\begin{equation}
    W_d\equiv\int_0^t\mathbf{F_d}\cdot \mathbf{v}dt'.
\end{equation}
The left-hand side (LHS) in equation \ref{eq:int} is the change in orbital energy, so
\begin{equation}
    \label{eq:result}
    \Delta E_\text{orb} =-G M_2 \int_0^t \frac{\dot{r}}{r}\frac{dM_\text{enc}}{dr}dt'+W_d.
\end{equation}
This equation states that the change in orbital energy has two sources: the change in the enclosed mass, and the work done by the drag forces. The work done by the drag forces is negative because the forces oppose the velocity of the companion. On the other hand, the first term on the RHS is positive if the orbit of the companion shrinks ($\dot{r}<0$), so it energetically opposes the drag forces. Drag forces must therefore do more work to shrink the orbit of the companion when the enclosed mass is a function of radius (and therefore time) than when the enclosed mass is constant. Equivalently, as the orbit of the companion shrinks, the companion is less bound to the new, smaller enclosed mass, than it would have been if the enclosed mass had remained constant. Therefore, additional work must be done to further shrink its orbit.

Using $\dot{r}dt'=dr'$, introducing the energy deposited into the envelope $E_\text{dep}\equiv-W_d$, and rearranging,
\begin{equation}
    \label{eq:revised_formalism}
    E_\text{dep}=-\Delta E_\text{orb} - \Delta E_m,
\end{equation}
where we defined
\begin{equation}
\label{eq:deltaem}
\Delta E_m = G M_2 \int_{r_0}^{r} \frac{1}{r'}\frac{dM_\text{enc}}{dr'}dr',
\end{equation}
where $r_0$ is the radial coordinate at $t=0$, and the subscript indicates this change of energy is a result of the changing enclosed mass. In appendix \ref{sec:appendix:validation}, we verify equation \ref{eq:revised_formalism} by numerically integrating equation \ref{eq:eom}.

%% file: comparison.tex
\section{Comparison to the standard formalism}\label{sec:comparison}
In this section, we assume for simplicity that the orbit remains approximately circular during CEE. In this approximation, the change in orbital energy is
\begin{equation}\label{eq:eorb}
    \Delta E_\text{orb} = -\frac{G M_2 M_\text{enc}}{2r} + \frac{G M_2 M_\text{enc,0}}{2r_0},
\end{equation}
where $M_\text{enc,0}$ is the enclosed mass at $t=0$. The rate of change of the orbital energy with radius is
\begin{equation}\label{eq:deorbdr}
    \frac{dE_\text{orb}}{dr} = \frac{G M_2}{2}\lp\frac{M_\text{enc}}{r^2}- \frac{1}{r}\frac{dM_\text{enc}}{dr}\rp.
\end{equation}
The radial rate of change of the additional term in the revised formalism is
\begin{equation}
    \label{eq:demdr}
    \frac{dE_m}{dr} = \frac{G M_2}{r}\frac{d\menc}{dr},
\end{equation}
and the radial rate of change of the energy deposition is
\begin{equation}
    \begin{split}
    \label{eq:dedepdr}
    \frac{dE_\text{dep}}{dr} = &-\frac{d\eorb}{dr}-\frac{dE_m}{dr}\\
    =&- \frac{G M_2}{2}\lp\frac{M_\text{enc}}{r^2} + \frac{1}{r}\frac{dM_\text{enc}}{dr}\rp.
    \end{split}
\end{equation}

\begin{figure*}[th!]
\centering
\includegraphics[width=0.495\textwidth]{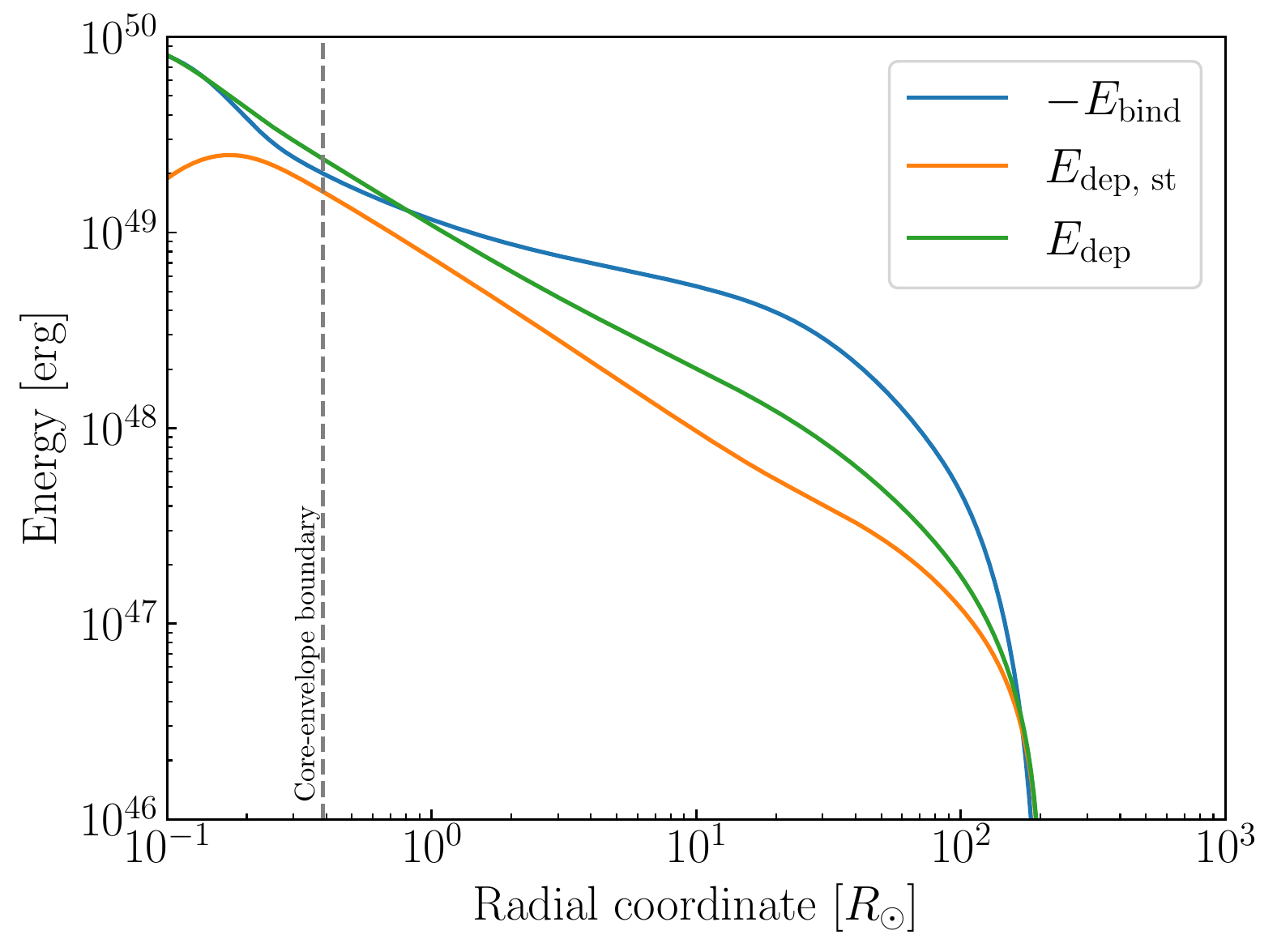}
\includegraphics[width=0.495\textwidth]{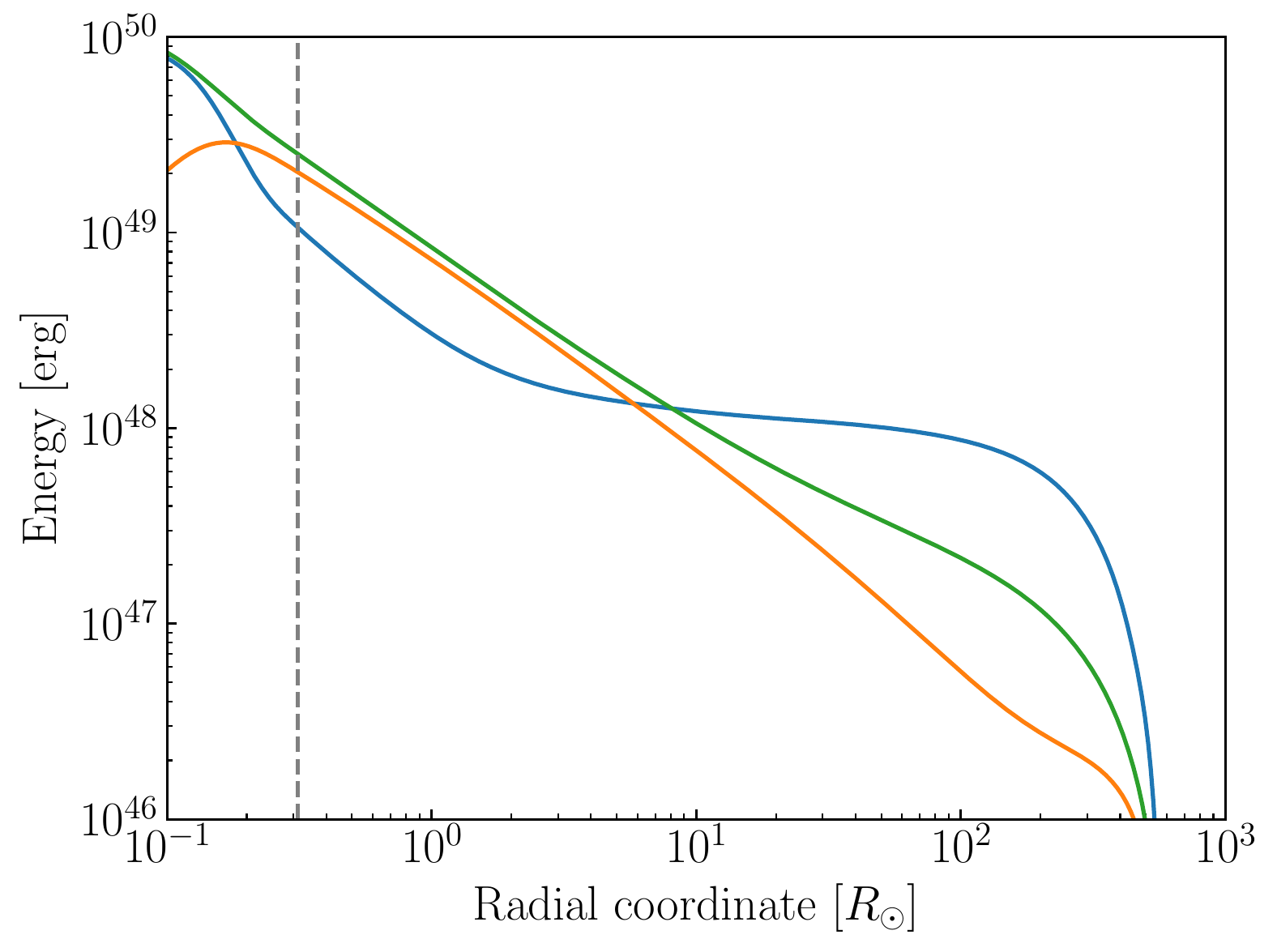}
\caption{Energy profiles as a function of radius for a $1.33M_\odot$ neutron star companion, as a function of the radial coordinate inside a $12M_\odot$ star evolved to $200R_\odot$ (left; first point in time at which envelope ejection is possible) and $500R_\odot$ (right). Lines in different colors show the magnitude of the binding energy $E_\text{bind}$ and the energy deposition predicted by the standard and revised formalisms ($E_\text{dep, st}$ and $E_\text{dep}$, respectively). The dashed vertical line shows the core-envelope boundary.}
\label{fig:ns:profiles}
\end{figure*}

\subsection{Inwards-increasing energy deposition}\label{sec:comparison:inwards}
The energy formalism sets the energy deposited into the envelope equal to $-\Delta E_\text{orb}$ (equation \ref{eq:eorb}). This quantity is assumed to be positive, which occurs only when
\begin{equation}
    \label{eq:inequality_standard_positive}
    \frac{M_\text{enc}}{M_\text{enc,0}}>\frac{r}{r_0};
\end{equation}
a positive energy deposition is not guaranteed for arbitrary stellar structure. While the inequality above typically holds above the core-envelope boundary in evolved stars, the dependence of the sign of the energy deposition on stellar structure is conceptually problematic. In the inner regions of the star, the rate of change of the orbital energy approaches zero and even reverses sign \citep[e.g. Figure \ref{fig:ns:profiles};][]{LawSmith2020,Moreno2021}, implying that a companion must gain energy to shrink its orbit further. This issue is particularly important for CEE mergers involving a compact object companion, whose orbital decay will continue until it reaches the center of the primary \citep[for an example of such a system, see][]{HutchinsonSmith2022}.

For the change in orbital energy to make an inwards-increasing contribution to the energy deposition, $d\eorb/dr$ (equation \ref{eq:deorbdr}) must be positive, so that the orbital energy decreases inwards. The contribution to the energy deposition rate from the additional term in the revised formalism (equation \ref{eq:demdr}) is positive and larger in magnitude than the negative term in equation \ref{eq:deorbdr}. The radial rate of change of the energy deposition (equation \ref{eq:dedepdr}) is therefore always negative in the revised formalism, so the deposited energy increases inwards. Since the energy deposition is initially zero, equation \ref{eq:dedepdr} also implies a positive energy deposition throughout CEE. This argument highlights an important qualitative property of the revised formalism, in alignment with physical intuition---the companion must always deposit energy into the envelope for its orbit to shrink, regardless of the structure of the primary.

\subsection{Relative term importance}
Both terms in the RHS of equation \ref{eq:revised_formalism} are proportional to the mass of the companion, so the relative importance between them is a function of stellar structure only.

The additional term will dominate the energy deposition when $\Delta E_m<\Delta E_\text{orb}$, which can be written as
\begin{equation}
\label{eq:term_importance_inequality}
\int_r^{r_0} \frac{1}{r'}\frac{dM_\text{enc}}{dr'}dr' > \frac{\menc}{2r}-\frac{\menc}{2r_0}.
\end{equation}
We will simplify this inequality by constructing a lower bound for the integral on the LHS. This integral satisfies
\begin{equation}
\label{eq:em_lower_limit}
\int_r^{r_0} \frac{1}{r'}\frac{dM_\text{enc}}{dr'}dr' > \int_r^{r_0} \frac{1}{r_0}\frac{dM_\text{enc}}{dr'}dr'= \frac{\menczero-\menc}{r_0}
\end{equation}
because $r'\leq r_0$ inside the integral. For the RHS of equation \ref{eq:em_lower_limit} to be larger than the RHS in equation \ref{eq:term_importance_inequality},
\begin{equation}
    \frac{\menc}{\menczero}<\frac{3}{2+r_0 / r}.
\end{equation}
A more general approximate criterion can be derived based on the expectation that the additional term will dominate the energy deposition in the parts of the star where its rate of change is consistently larger than that of the orbital energy, i.e.
\begin{equation}
    \frac{d E_m}{dr}>\frac{dE_\text{orb}}{dr},
\end{equation}
which reduces to
\begin{equation}
    \label{eq:profile_inequality}
    \frac{r}{\menc}\frac{d\menc}{dr}>\frac{1}{3}.
\end{equation}
This criterion is equivalent to the enclosed mass profile being locally steeper than a power law with exponent $1/3$.

\subsection{Common-envelope efficiency}
For the envelope to be ejected, the deposited energy must be equal to its binding energy. The fraction of this deposited energy that contributes to envelope ejection is referred to as the common-envelope efficiency $\alpha$ \citep{Livio1988}, and is defined in the standard formalism as
\begin{equation}\label{eq:efficiency_standard}
\alpha_\text{st}=-\frac{E_\text{bind,env}}{E_\text{dep,st}}=\frac{E_\text{bind, env}}{\Delta E_\text{orb}},
\end{equation}
where $E_\text{bind, env}$ is the binding energy of the envelope. If all sources of energy are included, the efficiency should be less than or equal to unity. However, the standard energy formalism can explain the formation of some systems, such as white dwarfs binaries \citep{Nelemans2000}, only if the efficiency is greater than unity. Some hydrodynamical simulations \citep[e.g.][]{LawSmith2020,Sand2020,Moreno2021,Lau2022} of CEE have measured this efficiency to be larger than unity. These results are typically attributed to the omission of other energy sources from equation \ref{eq:efficiency_standard}, such as the internal energy \citep{Han1994} and enthalpy \citep{Ivanova2011} of the envelope. The additional term $\Delta E_m$ behaves, in the standard definition of the common-envelope efficiency (equation \ref{eq:efficiency_standard}), as an additional unaccounted-for energy source, even though, like the $\Delta E_\text{orb}$ term, it is of gravitational origin. Values of the common-envelope efficiency above unity computed using equation \ref{eq:efficiency_standard} are therefore not necessarily unphysical, and do not necessarily arise only from the omission of other energy sources---they arise at least partly from an incomplete description of the gravitational energy deposition.

Equation \ref{eq:efficiency_standard} is frequently used to estimate the post-CEE orbital separation. An upper bound for this separation is obtained by setting $\alpha_\text{st}=1$. Since the revised formalism predicts a higher energy deposition, it also predicts a larger post-CEE separation, since the companion won't decay as far into the star before depositing enough energy to eject the envelope. 

%% file: methods.tex
\section{Numerical methods}
\label{sec:methods}
We construct stellar models using the modules for experiments in stellar astrophysics \citep[MESA,][]{Paxton2011}. We modified the \texttt{12M\_pre\_ms\_to\_core\_collapse} and \texttt{1M\_pre\_ms\_to\_wd} test suite parameter files to produce the stars shown in sections \ref{sec:discussion:ns} and \ref{sec:discussion:planet}, respectively. We set the metallicity to $Z=0.02$.

We estimate companions will eject the envelope of the primary if the deposited energy is greater in magnitude than the gravitational binding energy of the envelope. We define the core-envelope boundary as the location where the hydrogen mass fraction falls below one tenth.

We consider the CEE phase to end in a merger if the companion reaches the core-envelope boundary without ejecting the envelope or, in the case of compact companions, if their mass equals the enclosed mass, at which point they will tidally disrupt the enclosed mass. We estimate planetary companions will be destroyed, and unable to further deposit energy into the envelope, when (i) the kinetic energy per unit volume of the gas they encounter is equal to their binding energy per unit volume \citep{Jia2018}, or (ii) the enclosed mass tidally disrupts them, which will occur roughly when $r=R_2\lp M_\text{enc}/M_\text{2}\rp^{1/3}$, where $R_2$ is the radius of the companion. We compute the radius of planetary companions from their mass-radius relation \citep{Chabrier2009}, and their speed with respect to the surrounding medium by assuming circular Keplerian motion around the enclosed mass.

In numerical calculations, we use the gravitational constant value from \cite{Tiesinga2021}, and the solar system constants from \cite{Prsa2016}.

%% file: ns_binaries.tex
\subsection{The formation of neutron star binaries}\label{sec:discussion:ns}

\begin{figure}[t!]
    \centering
    \includegraphics[width=\columnwidth]{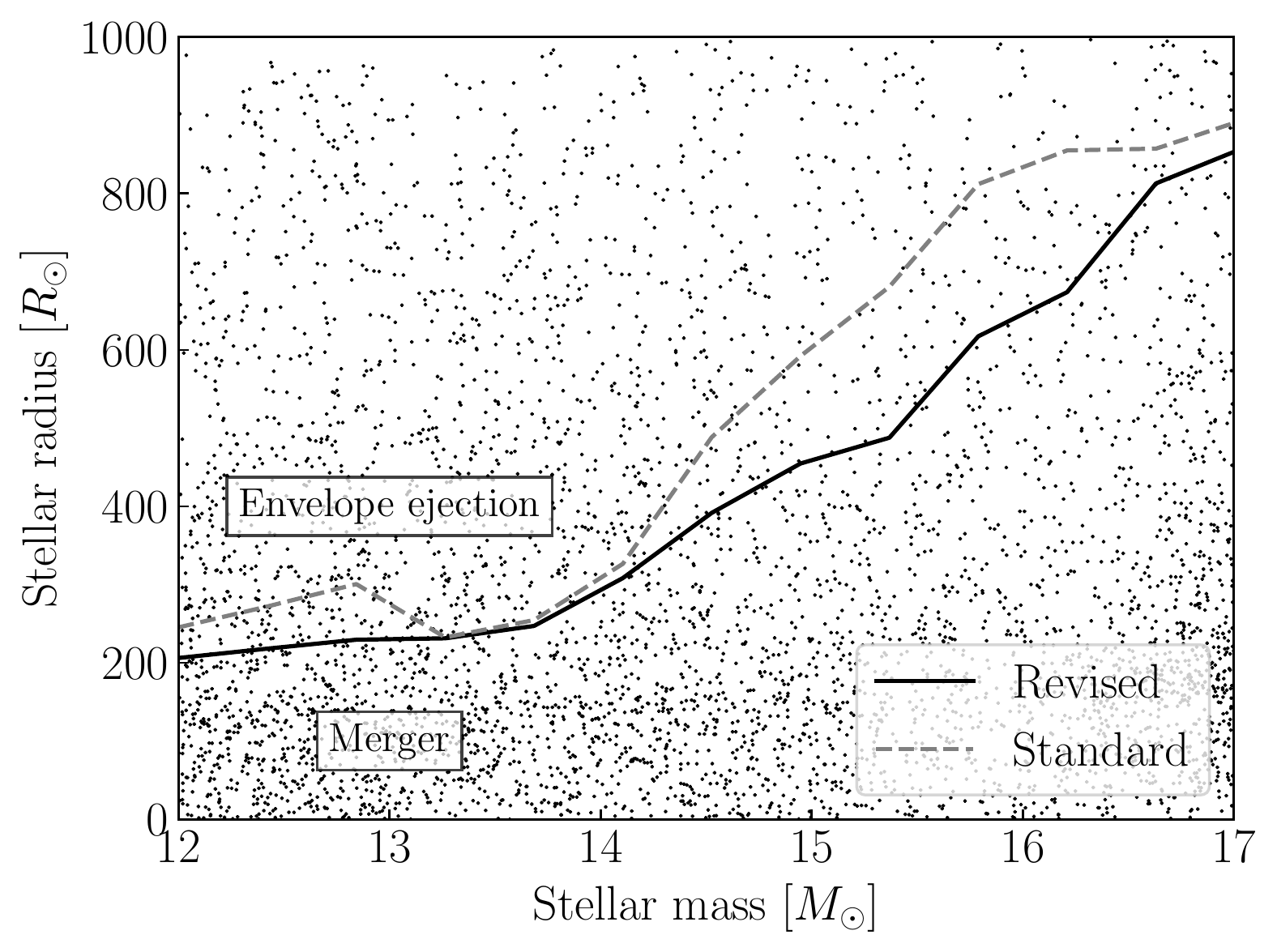}
    \caption{Minimum stellar radius at which a $1.33M_\odot$ neutron star can eject the envelope, as a function of stellar mass. Dashed and solid lines represent, respectively, the standard and revised energy formalisms. Scatter points show systems from a representative simulation \citep{COMPAS2022} from the COMPAS binary population synthesis code \citep{Riley2022}. The number of systems that eject the envelope in the parameter space of this Figure is higher by 20\% in the revised formalism.}
    \label{fig:ns_pop}
\end{figure}

Figure \ref{fig:ns:profiles} shows the energetic profile of the CEE between a $12M_\odot$ star and a $1.33M_\odot$ neutron star. This neutron star mass is the peak mass of the neutron star binary mass distribution \citep{Ozel2012}. Each panel in the figure corresponds to a different stage of stellar evolution. When the star has evolved to $200R_\odot$, the standard energy formalism underestimates the energy deposition by a factor of $\approx2$ throughout CEE. Consequently, it overestimates the mass required to eject the envelope by the same factor. The panel on the right shows the same energy profiles, but for a star evolved to $600R_\odot$. The energy deposition rates predicted by both formalisms at the core-envelope boundary are different by only about $50\%$ at this evolutionary stage, but by a factor of $\approx5$ in the outer envelope.

Figure \ref{fig:ns_pop} shows the stellar radius at which a $1.33M_\odot$ neutron star can eject the envelope, as a function of stellar mass. The dashed and solid lines represent, respectively, the standard and revised energy formalisms. More massive stars must be larger in size for envelope ejection to be possible. Scatter points show binary systems from a binary population synthesis simulation \citep{COMPAS2022} from the COMPAS code \citep{Riley2022}. We assume that companions are engulfed when the stellar radius equals the periapsis distance of their orbit. The number of systems that will lead to envelope ejection in the parameter space plotted in Figure \ref{fig:ns_pop} is $\approx20\%$ larger in the revised formalism.

Mass ejection during CEE will result in an infrared or optical transient. The properties of the plateau-shaped lightcurve depend on the ejecta mass $m_\text{ej}$ and speed at infinity $v_\text{ej}$ \citep{Ivanova2013}. The luminosity is proportional to $m_\text{ej}^{1/2}$ and $v_\text{ej}^{5/3}$, and the duration to $m_\text{ej}^{1/3}v_\text{ej}^{-1/3}$. Ejecta also radiates as a result of recombination with a luminosity proportional to its mass. The additional energy deposition in the outer envelope shown in Figure \ref{fig:ns:profiles} indicates higher ejecta mass and/or speed than predicted by the standard formalism. Since the transient luminosity increases with both ejecta mass and speed, the revised formalism predicts a more luminous plateau. Differences in plateau duration depend on how energy is distributed within the envelope, since the duration depends on ejecta mass and speed oppositely. For example, if there is no energy redistribution within the envelope, the same mass will be ejected, but at a significantly higher speed, resulting in a shorter plateau.

%% file: planetary_engulfment.tex
\subsection{Planetary engulfment}
\label{sec:discussion:planet}
\begin{figure}[t!]
    \centering
    \includegraphics[width=\columnwidth]{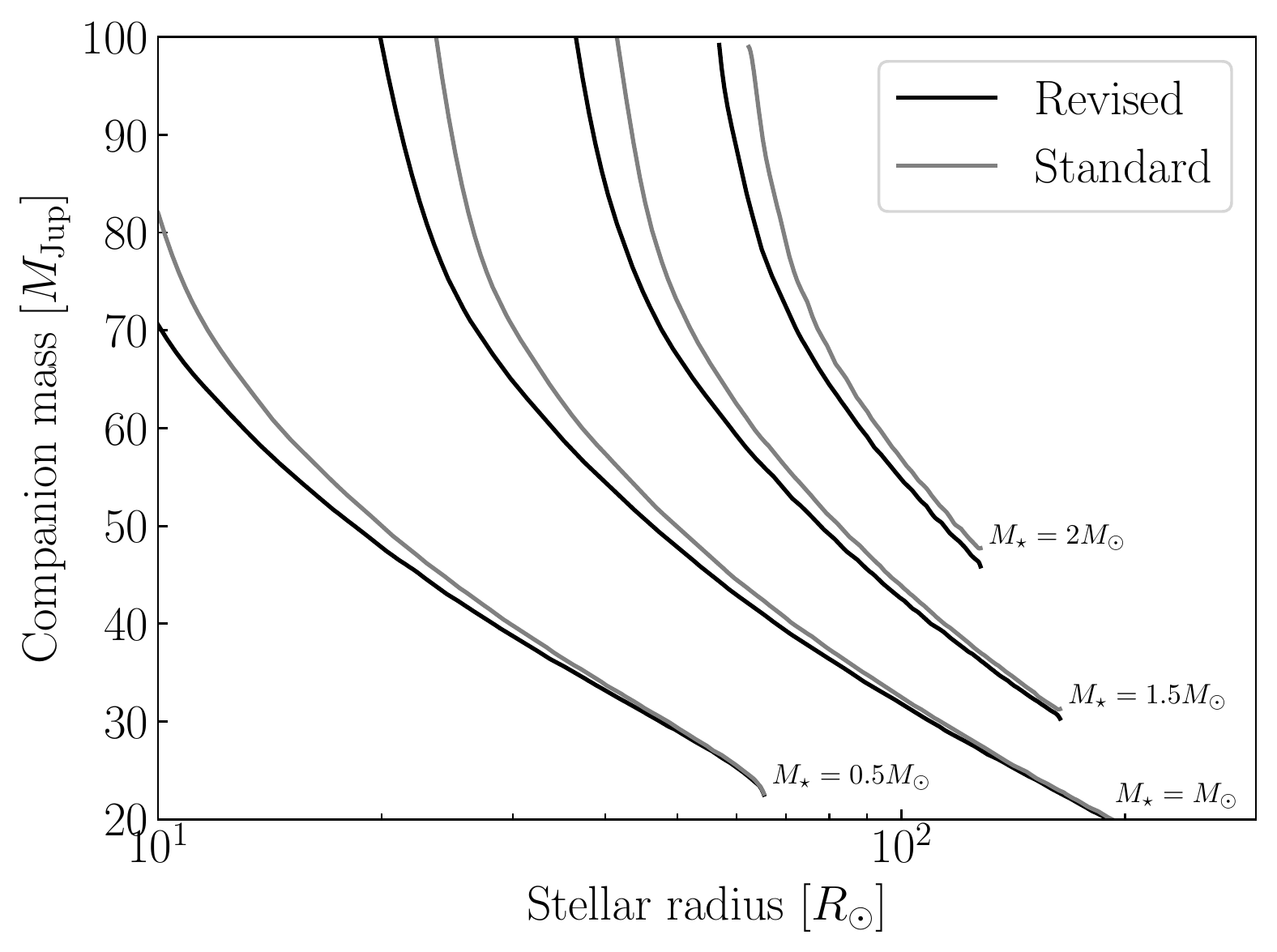}
    \caption{Minimum companion mass required to eject the envelope of stars of several masses, as a function of their radius. The dashed and solid lines correspond to the standard and revised energy formalisms, respectively.}
    \label{fig:minimum_mass_planet}
\end{figure}

Figure \ref{fig:minimum_mass_planet} shows the minimum companion mass required to eject the envelope of stars of different mass as a function of their radius. The standard formalism predicts larger companion masses than the revised formalism by about $25\%$ early in the red giant branch, but the differences are $\leq1\%$ near the tip of the red giant branch, when the star is more centrally concentrated.

\begin{figure*}[th!]
\centering
\includegraphics[width=0.495\textwidth]{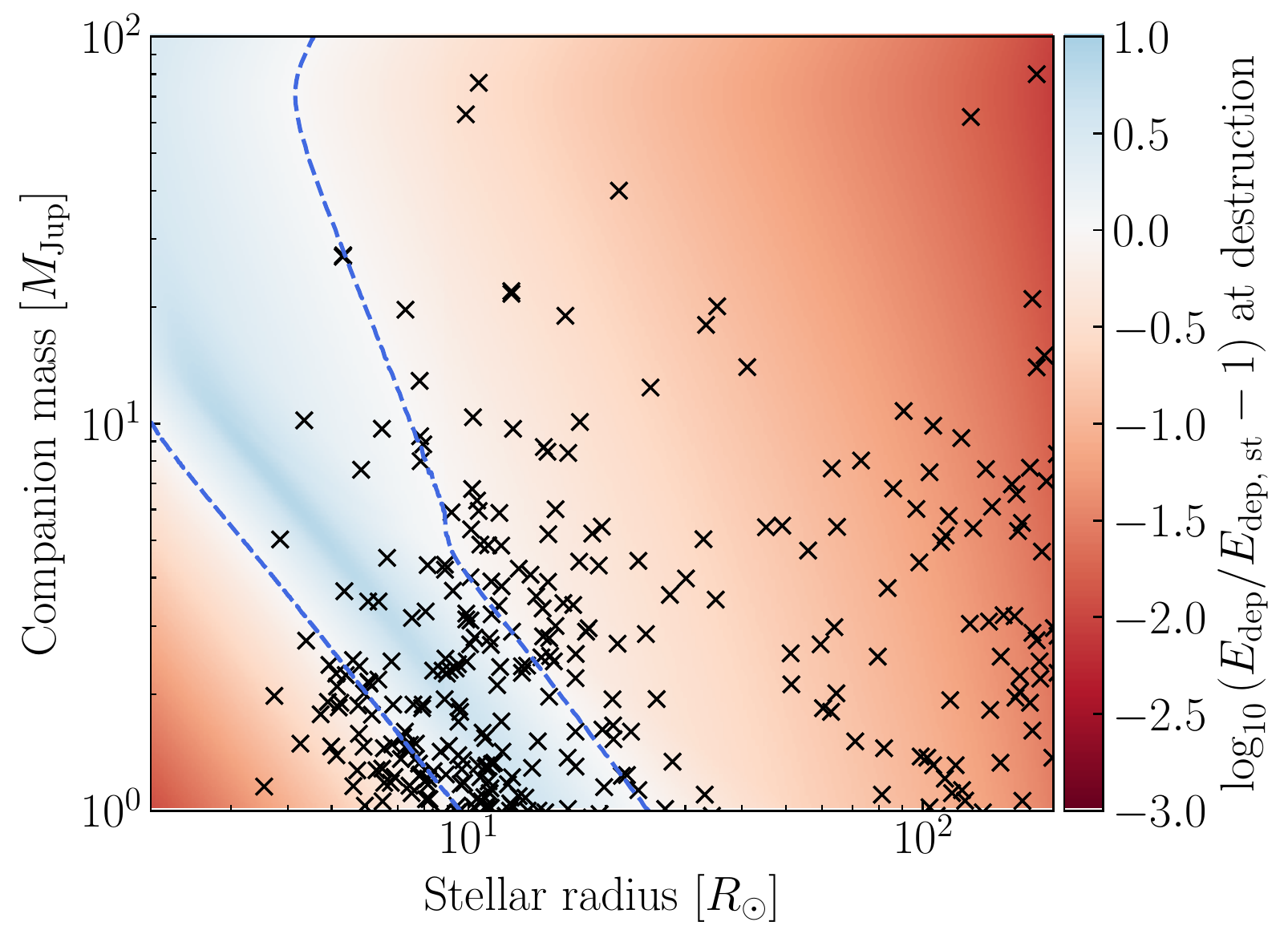}
\includegraphics[width=0.495\textwidth]{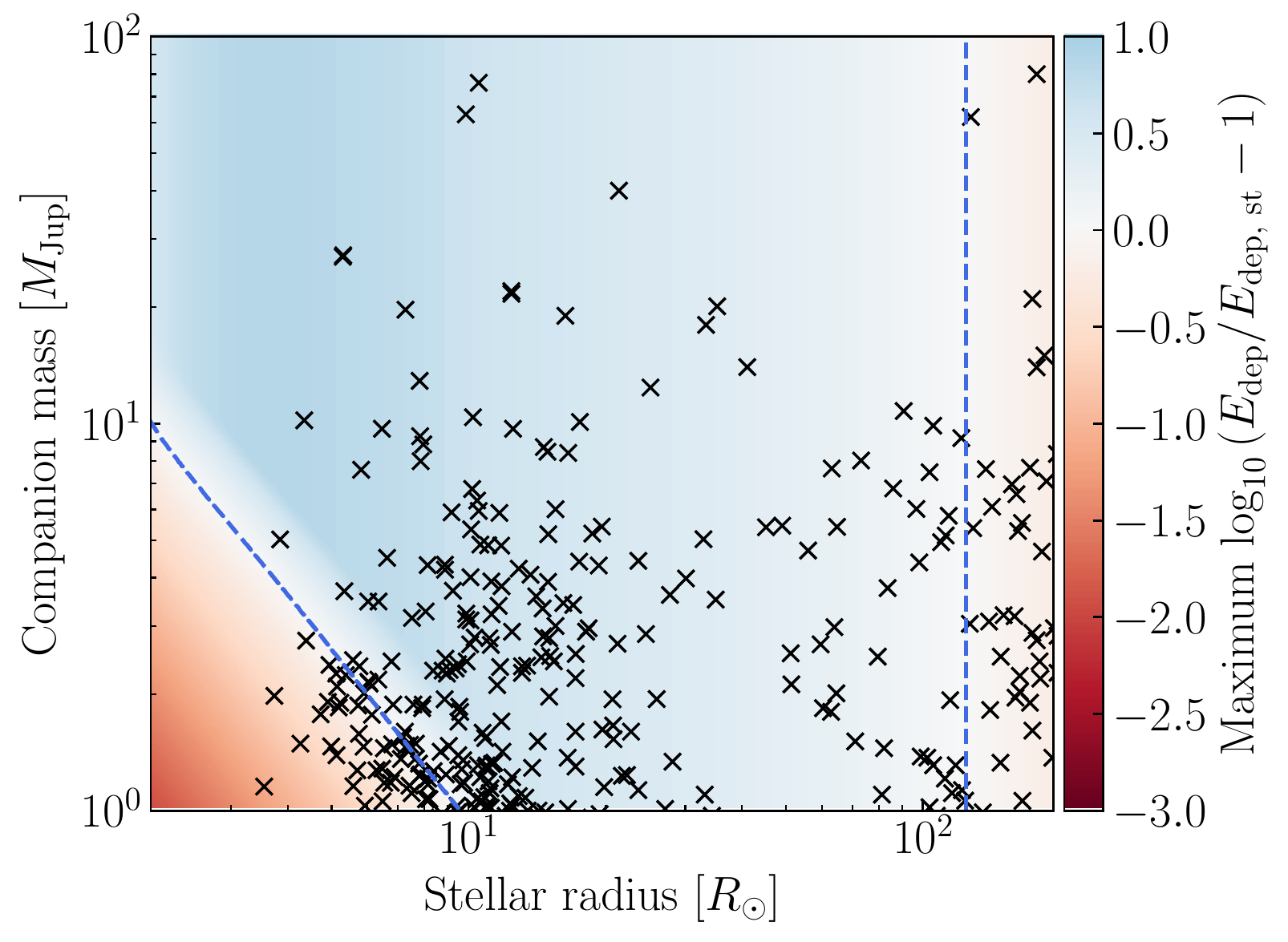}
\caption{Ratio between energy deposition predicted by the revised and standard energy formalisms, as a function of the stellar radius of a $1M_\odot$ star at the onset of common-envelope evolution and of companion mass. The left panel shows the ratio at the approximate location in the star where the companion will be destroyed, and the plot on the right shows the maximum ratio throughout its orbital decay. Scatter points show the known exoplanets \citep{exoplanetarchive}, assuming they are engulfed at their current orbital separations by $1M_\odot$ stars.}
\label{fig:planets:ratio}
\end{figure*}

The left panel in Figure \ref{fig:planets:ratio} shows the fractional difference between the energy deposition predicted by both formalisms at the location where the companion is destroyed, as a function of stellar radius at the onset of engulfment, and of companion mass. The energy deposition at that location is roughly proportional to the peak transient luminosity \citep{MacLeod2018}. Scatter plots show the known exoplanets, assuming they are engulfed at their current orbital separations by a $1M_\odot$ star. Contour lines show delimit the region where the relative difference is larger than unity.

Even if the total energy deposition at the point of destruction is similar, for most exoplanets the energy deposition will be larger than predicted by the standard formalism by a factor $\geq2$. The right panel in Figure \ref{fig:planets:ratio} shows the maximum fractional difference through engulfment (from onset to destruction) between the energy deposition predicted by both formalisms. As in the left panel, the dashed lines delimit the region where the relative difference is $\geq1$. Most exoplanets lie in this region, so at some point their energy deposition will be significantly different than predicted by the standard formalism. At large stellar radii this fractional difference is a function of stellar radius only. The tenuous envelope of the more evolved star allows for most companions to survive deeper than the point at which the difference between the two formalisms is at a maximum. When the star is more compact, companions above a critical mass will survive deeper than this maximum point, but for companions destroyed above this location the maximum fractional difference depends on their mass.

%% file: conclusion.tex
\section{Summary and conclusions}\label{sec:conclusion}

We derived an energy formalism for common-envelope evolution (CEE), starting from the equation of motion of the engulfed companion in the limit in which its mass is much smaller than the mass enclosed by its orbit. This formalism contains an additional term, compared to the standard formalism, that accounts for the energetic effect of the time-dependence of the mass enclosed by the orbit of the companion. Equations \ref{eq:revised_formalism} and \ref{eq:deltaem} quantitatively describe this revised formalism. Some of our key results are:

\begin{itemize}
  \item The revised formalism always predicts a higher energy deposition than the standard one, and it guarantees that the energy deposition is positive and inwards-increasing, regardless of the internal structure of the engulfing star.
  \item Differences in energy deposition are largest when the enclosed mass profile is the steepest. The energy deposition is higher by a factor $\gtrsim2$ in parts of the star where the local power law index of the enclosed mass profile is $\geq1/3$ (equation \ref{eq:profile_inequality}).
  \item Since the energy deposition is larger than predicted by the standard energy formalism, values above unity for the common-envelope efficiency as typically defined in the literature (equation \ref{eq:efficiency_standard}) are not necessarily unphysical, and might arise at least partly from an incomplete description of the gravitational energy deposition. Conclusions drawn from the assumption that the efficiency must be $\leq1$ should be reevaluated.
  \item For CEE involving a neutron star companion, the stellar radius at which envelope ejection becomes possible is smaller by up to $30\%$ (Figure \ref{fig:ns_pop}). At some stages of stellar evolution, the energy deposition in the outer envelope is a factor of $\approx5$ higher (Figure \ref{fig:ns:profiles}), thus changing the transient properties.
  \item For CEE involving a substellar companion, the mass required to eject the envelope of an evolved $1M_\odot$ star is lower by up to $\approx25\%$ (Figure \ref{fig:minimum_mass_planet}), whereas the energy deposition in the outer envelope is larger by a factor of up to $\approx7$ (Figure \ref{fig:planets:ratio}, left panel). Even in cases in which the peak energy deposition is similar, the energy deposition for most known substellar companions will at some point during engulfment be larger by a factor $\geq2$ (Figure \ref{fig:planets:ratio}, right panel).
\end{itemize}

The revised energy formalism we introduced in this Letter is valid in the limit in which the mass of the companion is small compared to the mass enclosed by its orbit. However, the progenitors of many systems formed through common-envelope evolution have mass ratios closer to unity (e.g., double white dwarfs). In future work, we will extend the revised formalism to study the energetics of these systems.

Presently, only hydrodynamical simulations can capture the three-dimensional redistribution of energy and mass during common-envelope evolution. The goal of this Letter is to improve on an analytical framework useful to validate and interpret the results of these simulations. This synergy between simplified analytical frameworks and computational models is crucial to understanding the dynamics and energetics of common-envelope evolution.

%% file: appendix.tex
\appendix
\section{Numerical validation}\label{sec:appendix:validation}
We verify equation \ref{eq:revised_formalism} by integrating the equation of motion of the companion (equation \ref{eq:eom}) numerically using the code described in \cite{Yarza2022}. We initially place the companion in a circular orbit at a separation $r=0.9R_\star$, where $R_\star$ is the radius of the star. The results of Section \ref{sec:derivation} are independent of the form of the drag $\mathbf{F_d}$; in our numerical integration we consider gravitational drag, given by
\begin{equation}
F_g=- \pi R_a^2\rho v^2\mathbf{\hat{v}},
\end{equation}
where $R_a=2G M_2/v^2$ is the accretion radius of the companion and $\mathbf{\hat{v}}$ is the unit vector in the direction of its velocity. The energy deposition from equation \ref{eq:revised_formalism} matches the numerical energy deposition for the systems in Figure \ref{fig:ns:profiles} to within a fractional difference of $3.7\times10^{-4}$. The leading source of error is likely energy conservation in the numerical integrator.

Three-dimensional hydrodynamical simulations have found that the work done by the drag forces and the change in orbital energy are not equal \citep[e.g., Figure 2 of][]{Wu2020}.

\section{Fitting formul\AE{}}\label{sec:appendix:fitting_formulae}
We provide fitting formul\ae{} for some results of this letter. See Appendix \ref{sec:appendix:software} for the data these formul\ae{} fit. The stellar radius at which envelope ejection by a $1.3M_\odot$ neutron star is possible is given by
\begin{equation}
\frac{R_{\star,\text{ min}}}{R_\odot}=\lp 1 + a_1 x + a_2 x^2\rp^{-1}\sum_{i=0} b_i x^i,
\end{equation}
where $x \equiv 0.1979\lp M_\star/M_\odot - 12\rp$, $a_1=-3.9849$, $a_2=4.5065$, and
\begin{equation}
\begin{split}
b_i=&\lb206.1,-717.32,1628.3,-11902,51307,\right.\\&\left.-99586,92123,-31739\rb.
\end{split}
\end{equation}
This formula is valid in the stellar mass range $12\leq M/M_\odot\leq17$, with a relative error $\leq4.4\%$ compared to Figure \ref{fig:ns_pop}.

The companion mass required to eject the envelope of a star of between one half and two solar masses as a function of its radius (Figure \ref{fig:minimum_mass_planet}) is given to within $6.2\%$ by
\begin{equation}
\frac{M_{2,\text{ min}}}{M_\text{Jup}}=\ls \sum_{i=0,j=0}a_{ij}x^i y^j \rs\ls \sum_{i=0,j=0}b_{ij}x^i y^j \rs^{-1},
\end{equation}
where
\begin{gather}
a_{ij}=\begin{pmatrix}
1290 & -199 & -0.879\\
-3420 & 748 & 0.29
\end{pmatrix},\\
b_{ij}=\begin{pmatrix}
0 & -5.25 & 0.0103\\
67.3 & 17.6 & 0.0624\\
-190 & -3.14 & -0.00601
\end{pmatrix},
\end{gather}
$x\equiv\lp M_\star/M_\odot\rp$, and $y\equiv\lp R_\star/R_\odot\rp$. This formula is valid between the radius at which envelope by companions with masses $\leq100M_\text{Jup}$ becomes possible,
\begin{equation}
\frac{R_\star}{R_\odot}=-1.94 + 14.58x + 7.4 x^2
\end{equation}
(to within $1.2\%$), and the radius at the tip of the red giant branch (to within $1.8\%$),
\begin{equation}
\frac{R_\star}{R_\odot}=\lp1-2.1359x+1.2322x^2\rp^{-1}\sum_{i=0}c_i x^i,
\end{equation}
where
\begin{equation}
c_i=\lb-81.05,494.27,-870.65,603.71,-127.44\rb.
\end{equation}

\section{Data and software availability}\label{sec:appendix:software}
The software and data required to reproduce the results of this letter are available under the digital object identifier (DOI) \href{https://doi.org/10.5281/zenodo.7058241}{10.5281/zenodo.7058241}. This DOI also contains code to compute the energy deposition according to the revised formalism (equations \ref{eq:revised_formalism} and \ref{eq:deltaem}).

%% file: main.bbl
\begin{thebibliography}{}
\expandafter\ifx\csname natexlab\endcsname\relax\def\natexlab#1{#1}\fi
\providecommand{\url}[1]{\href{#1}{#1}}
\providecommand{\dodoi}[1]{doi:~\href{http://doi.org/#1}{\nolinkurl{#1}}}
\providecommand{\doeprint}[1]{\href{http://ascl.net/#1}{\nolinkurl{http://ascl.net/#1}}}
\providecommand{\doarXiv}[1]{\href{https://arxiv.org/abs/#1}{\nolinkurl{https://arxiv.org/abs/#1}}}

\bibitem[{{Aguilera-G{\'o}mez}
  {et~al.}(2016{\natexlab{a}}){Aguilera-G{\'o}mez}, {Chanam{\'e}},
  {Pinsonneault}, \& {Carlberg}}]{AguileraGomez2016}
{Aguilera-G{\'o}mez}, C., {Chanam{\'e}}, J., {Pinsonneault}, M.~H., \&
  {Carlberg}, J.~K. 2016{\natexlab{a}}, \apjl, 833, L24,
  \dodoi{10.3847/2041-8213/833/2/L24}

\bibitem[{{Aguilera-G{\'o}mez}
  {et~al.}(2016{\natexlab{b}}){Aguilera-G{\'o}mez}, {Chanam{\'e}},
  {Pinsonneault}, \& {Carlberg}}]{AguileraGomez2016a}
---. 2016{\natexlab{b}}, \apj, 829, 127, \dodoi{10.3847/0004-637X/829/2/127}

\bibitem[{{Angulo} {et~al.}(1999){Angulo}, {Arnould}, {Rayet}, {Descouvemont},
  {Baye}, {Leclercq-Willain}, {Coc}, {Barhoumi}, {Aguer}, {Rolfs}, {Kunz},
  {Hammer}, {Mayer}, {Paradellis}, {Kossionides}, {Chronidou}, {Spyrou},
  {degl'Innocenti}, {Fiorentini}, {Ricci}, {Zavatarelli}, {Providencia},
  {Wolters}, {Soares}, {Grama}, {Rahighi}, {Shotter}, \& {Lamehi
  Rachti}}]{Angulo1999}
{Angulo}, C., {Arnould}, M., {Rayet}, M., {et~al.} 1999, \nphysa, 656, 3,
  \dodoi{10.1016/S0375-9474(99)00030-5}

\bibitem[{{Bear} {et~al.}(2021){Bear}, {Merlov}, {Arad}, \& {Soker}}]{Bear2021}
{Bear}, E., {Merlov}, A., {Arad}, Y., \& {Soker}, N. 2021, \mnras, 507, 414,
  \dodoi{10.1093/mnras/stab2149}

\bibitem[{{Blouin} {et~al.}(2020){Blouin}, {Shaffer}, {Saumon}, \&
  {Starrett}}]{Blouin2020}
{Blouin}, S., {Shaffer}, N.~R., {Saumon}, D., \& {Starrett}, C.~E. 2020, \apj,
  899, 46, \dodoi{10.3847/1538-4357/ab9e75}

\bibitem[{{Cassisi} {et~al.}(2007){Cassisi}, {Potekhin}, {Pietrinferni},
  {Catelan}, \& {Salaris}}]{Cassisi2007}
{Cassisi}, S., {Potekhin}, A.~Y., {Pietrinferni}, A., {Catelan}, M., \&
  {Salaris}, M. 2007, \apj, 661, 1094, \dodoi{10.1086/516819}

\bibitem[{{Chabrier} {et~al.}(2009){Chabrier}, {Baraffe}, {Leconte},
  {Gallardo}, \& {Barman}}]{Chabrier2009}
{Chabrier}, G., {Baraffe}, I., {Leconte}, J., {Gallardo}, J., \& {Barman}, T.
  2009, in American Institute of Physics Conference Series, Vol. 1094, 15th
  Cambridge Workshop on Cool Stars, Stellar Systems, and the Sun, ed.
  E.~{Stempels}, 102--111, \dodoi{10.1063/1.3099078}

\bibitem[{{Chugunov} {et~al.}(2007){Chugunov}, {Dewitt}, \&
  {Yakovlev}}]{Chugunov2007}
{Chugunov}, A.~I., {Dewitt}, H.~E., \& {Yakovlev}, D.~G. 2007, \prd, 76,
  025028, \dodoi{10.1103/PhysRevD.76.025028}

\bibitem[{{Cyburt} {et~al.}(2010){Cyburt}, {Amthor}, {Ferguson}, {Meisel},
  {Smith}, {Warren}, {Heger}, {Hoffman}, {Rauscher}, {Sakharuk}, {Schatz},
  {Thielemann}, \& {Wiescher}}]{Cyburt2010}
{Cyburt}, R.~H., {Amthor}, A.~M., {Ferguson}, R., {et~al.} 2010, \apjs, 189,
  240, \dodoi{10.1088/0067-0049/189/1/240}

\bibitem[{{Ferguson} {et~al.}(2005){Ferguson}, {Alexander}, {Allard}, {Barman},
  {Bodnarik}, {Hauschildt}, {Heffner-Wong}, \& {Tamanai}}]{Ferguson2005}
{Ferguson}, J.~W., {Alexander}, D.~R., {Allard}, F., {et~al.} 2005, \apj, 623,
  585, \dodoi{10.1086/428642}

\bibitem[{{Fuller} {et~al.}(1985){Fuller}, {Fowler}, \& {Newman}}]{Fuller1985}
{Fuller}, G.~M., {Fowler}, W.~A., \& {Newman}, M.~J. 1985, \apj, 293, 1,
  \dodoi{10.1086/163208}

\bibitem[{Galassi {et~al.}(2009)}]{gsl}
Galassi, M., {et~al.} 2009, GNU Scientific Library Reference Manual, 3rd edn.

\bibitem[{{Han} {et~al.}(1994){Han}, {Podsiadlowski}, \& {Eggleton}}]{Han1994}
{Han}, Z., {Podsiadlowski}, P., \& {Eggleton}, P.~P. 1994, \mnras, 270, 121,
  \dodoi{10.1093/mnras/270.1.121}

\bibitem[{Harris {et~al.}(2020)Harris, Millman, van~der Walt, Gommers,
  Virtanen, Cournapeau, Wieser, Taylor, Berg, Smith, Kern, Picus, Hoyer, van
  Kerkwijk, Brett, Haldane, Fernández~del Río, Wiebe, Peterson,
  Gérard-Marchant, Sheppard, Reddy, Weckesser, Abbasi, Gohlke, \&
  Oliphant}]{Harris2020}
Harris, C.~R., Millman, K.~J., van~der Walt, S.~J., {et~al.} 2020, Nature, 585,
  357–362, \dodoi{10.1038/s41586-020-2649-2}

\bibitem[{Hunter(2007)}]{matplotlib}
Hunter, J.~D. 2007, Computing in Science \& Engineering, 9, 90,
  \dodoi{10.1109/MCSE.2007.55}

\bibitem[{{Hutchinson-Smith} {et~al.}(2022){Hutchinson-Smith}, {Wallace
  Everson}, \& {Ramirez-Ruiz}}]{HutchinsonSmith2022}
{Hutchinson-Smith}, T., {Wallace Everson}, R., \& {Ramirez-Ruiz}, E. 2022, in
  preparation

\bibitem[{{Iglesias} \& {Rogers}(1993)}]{Iglesias1993}
{Iglesias}, C.~A., \& {Rogers}, F.~J. 1993, \apj, 412, 752,
  \dodoi{10.1086/172958}

\bibitem[{{Iglesias} \& {Rogers}(1996)}]{Iglesias1996}
---. 1996, \apj, 464, 943, \dodoi{10.1086/177381}

\bibitem[{{Irwin}(2004)}]{Irwin2004}
{Irwin}, A.~W. 2004, The FreeEOS Code for Calculating the Equation of State for
  Stellar Interiors.
\newblock \url{http://freeeos.sourceforge.net/}

\bibitem[{{Itoh} {et~al.}(1996){Itoh}, {Hayashi}, {Nishikawa}, \&
  {Kohyama}}]{Itoh1996}
{Itoh}, N., {Hayashi}, H., {Nishikawa}, A., \& {Kohyama}, Y. 1996, \apjs, 102,
  411, \dodoi{10.1086/192264}

\bibitem[{{Ivanova} \& {Chaichenets}(2011)}]{Ivanova2011}
{Ivanova}, N., \& {Chaichenets}, S. 2011, \apjl, 731, L36,
  \dodoi{10.1088/2041-8205/731/2/L36}

\bibitem[{{Ivanova} {et~al.}(2013{\natexlab{a}}){Ivanova}, {Justham}, {Avendano
  Nandez}, \& {Lombardi}}]{Ivanova2013}
{Ivanova}, N., {Justham}, S., {Avendano Nandez}, J.~L., \& {Lombardi}, J.~C.
  2013{\natexlab{a}}, Science, 339, 433, \dodoi{10.1126/science.1225540}

\bibitem[{{Ivanova} {et~al.}(2013{\natexlab{b}}){Ivanova}, {Justham}, {Chen},
  {De Marco}, {Fryer}, {Gaburov}, {Ge}, {Glebbeek}, {Han}, {Li}, {Lu}, {Marsh},
  {Podsiadlowski}, {Potter}, {Soker}, {Taam}, {Tauris}, {van den Heuvel}, \&
  {Webbink}}]{Ivanova2013a}
{Ivanova}, N., {Justham}, S., {Chen}, X., {et~al.} 2013{\natexlab{b}}, \aapr,
  21, 59, \dodoi{10.1007/s00159-013-0059-2}

\bibitem[{{Jermyn} {et~al.}(2021){Jermyn}, {Schwab}, {Bauer}, {Timmes}, \&
  {Potekhin}}]{Jermyn2021}
{Jermyn}, A.~S., {Schwab}, J., {Bauer}, E., {Timmes}, F.~X., \& {Potekhin},
  A.~Y. 2021, \apj, 913, 72, \dodoi{10.3847/1538-4357/abf48e}

\bibitem[{{Jia} \& {Spruit}(2018)}]{Jia2018}
{Jia}, S., \& {Spruit}, H.~C. 2018, \apj, 864, 169,
  \dodoi{10.3847/1538-4357/aad77c}

\bibitem[{{Kramer} {et~al.}(2020){Kramer}, {Schneider}, {Ohlmann}, {Geier},
  {Schaffenroth}, {Pakmor}, \& {R{\"o}pke}}]{Kramer2020}
{Kramer}, M., {Schneider}, F.~R.~N., {Ohlmann}, S.~T., {et~al.} 2020, \aap,
  642, A97, \dodoi{10.1051/0004-6361/202038702}

\bibitem[{{Langanke} \& {Mart{\'\i}nez-Pinedo}(2000)}]{Langanke2000}
{Langanke}, K., \& {Mart{\'\i}nez-Pinedo}, G. 2000, \nphysa, 673, 481,
  \dodoi{10.1016/S0375-9474(00)00131-7}

\bibitem[{{Lau} {et~al.}(2022){Lau}, {Hirai}, {Gonz{\'a}lez-Bol{\'\i}var},
  {Price}, {De Marco}, \& {Mandel}}]{Lau2022}
{Lau}, M. Y.~M., {Hirai}, R., {Gonz{\'a}lez-Bol{\'\i}var}, M., {et~al.} 2022,
  \mnras, 512, 5462, \dodoi{10.1093/mnras/stac049}

\bibitem[{{Law-Smith} {et~al.}(2020){Law-Smith}, {Everson}, {Ramirez-Ruiz}, {de
  Mink}, {van Son}, {G{\"o}tberg}, {Zellmann}, {Vigna-G{\'o}mez}, {Renzo},
  {Wu}, {Schr{\o}der}, {Foley}, \& {Hutchinson-Smith}}]{LawSmith2020}
{Law-Smith}, J. A.~P., {Everson}, R.~W., {Ramirez-Ruiz}, E., {et~al.} 2020,
  arXiv e-prints, arXiv:2011.06630.
\newblock \doarXiv{2011.06630}

\bibitem[{{Livio} \& {Soker}(1984)}]{Livio1984}
{Livio}, M., \& {Soker}, N. 1984, \mnras, 208, 763,
  \dodoi{10.1093/mnras/208.4.763}

\bibitem[{{Livio} \& {Soker}(1988)}]{Livio1988}
---. 1988, \apj, 329, 764, \dodoi{10.1086/166419}

\bibitem[{{MacLeod} {et~al.}(2018){MacLeod}, {Cantiello}, \&
  {Soares-Furtado}}]{MacLeod2018}
{MacLeod}, M., {Cantiello}, M., \& {Soares-Furtado}, M. 2018, \apjl, 853, L1,
  \dodoi{10.3847/2041-8213/aaa5fa}

\bibitem[{{Moreno} {et~al.}(2021){Moreno}, {Schneider}, {Roepke}, {Ohlmann},
  {Pakmor}, {Podsiadlowski}, \& {Sand}}]{Moreno2021}
{Moreno}, M.~M., {Schneider}, F. R.~N., {Roepke}, F.~K., {et~al.} 2021, arXiv
  e-prints, arXiv:2111.12112.
\newblock \doarXiv{2111.12112}

\bibitem[{{NASA Exoplanet Archive}(2022)}]{exoplanetarchive}
{NASA Exoplanet Archive}. 2022, Planetary Systems Composite Parameters,
  Version: YYYY-MM-DD HH:MM,  NExScI-Caltech/IPAC, \dodoi{10.26133/NEA13}

\bibitem[{{Nelemans} \& {Tauris}(1998)}]{Nelemans1998}
{Nelemans}, G., \& {Tauris}, T.~M. 1998, \aap, 335, L85.
\newblock \doarXiv{astro-ph/9806011}

\bibitem[{{Nelemans} {et~al.}(2000){Nelemans}, {Verbunt}, {Yungelson}, \&
  {Portegies Zwart}}]{Nelemans2000}
{Nelemans}, G., {Verbunt}, F., {Yungelson}, L.~R., \& {Portegies Zwart}, S.~F.
  2000, \aap, 360, 1011.
\newblock \doarXiv{astro-ph/0006216}

\bibitem[{{Nordhaus} \& {Spiegel}(2013)}]{Nordhaus2013}
{Nordhaus}, J., \& {Spiegel}, D.~S. 2013, \mnras, 432, 500,
  \dodoi{10.1093/mnras/stt569}

\bibitem[{{Oda} {et~al.}(1994){Oda}, {Hino}, {Muto}, {Takahara}, \&
  {Sato}}]{Oda1994}
{Oda}, T., {Hino}, M., {Muto}, K., {Takahara}, M., \& {Sato}, K. 1994, Atomic
  Data and Nuclear Data Tables, 56, 231, \dodoi{10.1006/adnd.1994.1007}

\bibitem[{{{\"O}zel} {et~al.}(2012){{\"O}zel}, {Psaltis}, {Narayan}, \& {Santos
  Villarreal}}]{Ozel2012}
{{\"O}zel}, F., {Psaltis}, D., {Narayan}, R., \& {Santos Villarreal}, A. 2012,
  \apj, 757, 55, \dodoi{10.1088/0004-637X/757/1/55}

\bibitem[{{Paczynski}(1976)}]{Paczynski1976}
{Paczynski}, B. 1976, in Structure and Evolution of Close Binary Systems, ed.
  P.~{Eggleton}, S.~{Mitton}, \& J.~{Whelan}, Vol.~73, 75

\bibitem[{{Paxton} {et~al.}(2011){Paxton}, {Bildsten}, {Dotter}, {Herwig},
  {Lesaffre}, \& {Timmes}}]{Paxton2011}
{Paxton}, B., {Bildsten}, L., {Dotter}, A., {et~al.} 2011, \apjs, 192, 3,
  \dodoi{10.1088/0067-0049/192/1/3}

\bibitem[{{Paxton} {et~al.}(2013){Paxton}, {Cantiello}, {Arras}, {Bildsten},
  {Brown}, {Dotter}, {Mankovich}, {Montgomery}, {Stello}, {Timmes}, \&
  {Townsend}}]{Paxton2013}
{Paxton}, B., {Cantiello}, M., {Arras}, P., {et~al.} 2013, \apjs, 208, 4,
  \dodoi{10.1088/0067-0049/208/1/4}

\bibitem[{{Paxton} {et~al.}(2015){Paxton}, {Marchant}, {Schwab}, {Bauer},
  {Bildsten}, {Cantiello}, {Dessart}, {Farmer}, {Hu}, {Langer}, {Townsend},
  {Townsley}, \& {Timmes}}]{Paxton2015}
{Paxton}, B., {Marchant}, P., {Schwab}, J., {et~al.} 2015, \apjs, 220, 15,
  \dodoi{10.1088/0067-0049/220/1/15}

\bibitem[{{Paxton} {et~al.}(2018){Paxton}, {Schwab}, {Bauer}, {Bildsten},
  {Blinnikov}, {Duffell}, {Farmer}, {Goldberg}, {Marchant}, {Sorokina},
  {Thoul}, {Townsend}, \& {Timmes}}]{Paxton2018}
{Paxton}, B., {Schwab}, J., {Bauer}, E.~B., {et~al.} 2018, \apjs, 234, 34,
  \dodoi{10.3847/1538-4365/aaa5a8}

\bibitem[{{Paxton} {et~al.}(2019){Paxton}, {Smolec}, {Schwab}, {Gautschy},
  {Bildsten}, {Cantiello}, {Dotter}, {Farmer}, {Goldberg}, {Jermyn}, {Kanbur},
  {Marchant}, {Thoul}, {Townsend}, {Wolf}, {Zhang}, \& {Timmes}}]{Paxton2019}
{Paxton}, B., {Smolec}, R., {Schwab}, J., {et~al.} 2019, \apjs, 243, 10,
  \dodoi{10.3847/1538-4365/ab2241}

\bibitem[{{Potekhin} \& {Chabrier}(2010)}]{Potekhin2010}
{Potekhin}, A.~Y., \& {Chabrier}, G. 2010, Contributions to Plasma Physics, 50,
  82, \dodoi{10.1002/ctpp.201010017}

\bibitem[{{Poutanen}(2017)}]{Poutanen2017}
{Poutanen}, J. 2017, \apj, 835, 119, \dodoi{10.3847/1538-4357/835/2/119}

\bibitem[{{Pr{\v{s}}a} {et~al.}(2016){Pr{\v{s}}a}, {Harmanec}, {Torres},
  {Mamajek}, {Asplund}, {Capitaine}, {Christensen-Dalsgaard}, {Depagne},
  {Haberreiter}, {Hekker}, {Hilton}, {Kopp}, {Kostov}, {Kurtz}, {Laskar},
  {Mason}, {Milone}, {Montgomery}, {Richards}, {Schmutz}, {Schou}, \&
  {Stewart}}]{Prsa2016}
{Pr{\v{s}}a}, A., {Harmanec}, P., {Torres}, G., {et~al.} 2016, \aj, 152, 41,
  \dodoi{10.3847/0004-6256/152/2/41}

\bibitem[{{Riley} {et~al.}(2022){Riley}, {Agrawal}, {Barrett}, {Boyett},
  {Broekgaarden}, {Chattopadhyay}, {Gaebel}, {Gittins}, {Hirai}, {Howitt},
  {Justham}, {Khandelwal}, {Kummer}, {Lau}, {Mandel}, {de Mink}, {Neijssel},
  {Riley}, {van Son}, {Stevenson}, {Vigna-G{\'o}mez}, {Vinciguerra}, {Wagg},
  {Willcox}, \& {Team Compas}}]{Riley2022}
{Riley}, J., {Agrawal}, P., {Barrett}, J.~W., {et~al.} 2022, \apjs, 258, 34,
  \dodoi{10.3847/1538-4365/ac416c}

\bibitem[{{Rogers} \& {Nayfonov}(2002)}]{Rogers2002}
{Rogers}, F.~J., \& {Nayfonov}, A. 2002, \apj, 576, 1064,
  \dodoi{10.1086/341894}

\bibitem[{{Sand} {et~al.}(2020){Sand}, {Ohlmann}, {Schneider}, {Pakmor}, \&
  {R{\"o}pke}}]{Sand2020}
{Sand}, C., {Ohlmann}, S.~T., {Schneider}, F. R.~N., {Pakmor}, R., \&
  {R{\"o}pke}, F.~K. 2020, \aap, 644, A60, \dodoi{10.1051/0004-6361/202038992}

\bibitem[{{Sandquist} {et~al.}(1998){Sandquist}, {Taam}, {Lin}, \&
  {Burkert}}]{Sandquist1998}
{Sandquist}, E., {Taam}, R.~E., {Lin}, D.~N.~C., \& {Burkert}, A. 1998, \apjl,
  506, L65, \dodoi{10.1086/311633}

\bibitem[{{Sandquist} {et~al.}(2002){Sandquist}, {Dokter}, {Lin}, \&
  {Mardling}}]{Sandquist2002}
{Sandquist}, E.~L., {Dokter}, J.~J., {Lin}, D.~N.~C., \& {Mardling}, R.~A.
  2002, \apj, 572, 1012, \dodoi{10.1086/340452}

\bibitem[{{Saumon} {et~al.}(1995){Saumon}, {Chabrier}, \& {van
  Horn}}]{Saumon1995}
{Saumon}, D., {Chabrier}, G., \& {van Horn}, H.~M. 1995, \apjs, 99, 713,
  \dodoi{10.1086/192204}

\bibitem[{{Siess} \& {Livio}(1999)}]{Siess1999}
{Siess}, L., \& {Livio}, M. 1999, \mnras, 308, 1133,
  \dodoi{10.1046/j.1365-8711.1999.02784.x}

\bibitem[{{Soares-Furtado} {et~al.}(2021){Soares-Furtado}, {Cantiello},
  {MacLeod}, \& {Ness}}]{SoaresFurtado2021}
{Soares-Furtado}, M., {Cantiello}, M., {MacLeod}, M., \& {Ness}, M.~K. 2021,
  \aj, 162, 273, \dodoi{10.3847/1538-3881/ac273c}

\bibitem[{{Staff} {et~al.}(2016){Staff}, {De Marco}, {Wood}, {Galaviz}, \&
  {Passy}}]{Staff2016}
{Staff}, J.~E., {De Marco}, O., {Wood}, P., {Galaviz}, P., \& {Passy}, J.-C.
  2016, \mnras, 458, 832, \dodoi{10.1093/mnras/stw331}

\bibitem[{{Team COMPAS}(2022)}]{COMPAS2022}
{Team COMPAS}. 2022, {COMPAS v02.27.05 output at representative metallicities},
  02.27.05,  Zenodo, \dodoi{10.5281/zenodo.6346444}

\bibitem[{{The HDF Group}(1997-NNNN)}]{hdf5}
{The HDF Group}. 1997-NNNN, {Hierarchical Data Format, version 5}

\bibitem[{Tiesinga {et~al.}(2021)Tiesinga, Mohr, Newell, \&
  Taylor}]{Tiesinga2021}
Tiesinga, E., Mohr, P.~J., Newell, D.~B., \& Taylor, B.~N. 2021, Rev. Mod.
  Phys., 93, 025010, \dodoi{10.1103/RevModPhys.93.025010}

\bibitem[{{Timmes} \& {Swesty}(2000)}]{Timmes2000}
{Timmes}, F.~X., \& {Swesty}, F.~D. 2000, \apjs, 126, 501,
  \dodoi{10.1086/313304}

\bibitem[{Townsend(2021)}]{Townsend2021}
Townsend, R. 2021, MESA SDK for Linux, 21.4.1,  Zenodo,
  \dodoi{10.5281/zenodo.5802444}

\bibitem[{{van den Heuvel}(1976)}]{vandenHeuvel1976}
{van den Heuvel}, E.~P.~J. 1976, in Structure and Evolution of Close Binary
  Systems, ed. P.~{Eggleton}, S.~{Mitton}, \& J.~{Whelan}, Vol.~73, 35

\bibitem[{Virtanen {et~al.}(2020)Virtanen, Gommers, Oliphant, Haberland, Reddy,
  Cournapeau, Burovski, Peterson, Weckesser, Bright, {van der Walt}, Brett,
  Wilson, Millman, Mayorov, Nelson, Jones, Kern, Larson, Carey, Polat, Feng,
  Moore, {VanderPlas}, Laxalde, Perktold, Cimrman, Henriksen, Quintero, Harris,
  Archibald, Ribeiro, Pedregosa, {van Mulbregt}, \& {SciPy 1.0
  Contributors}}]{scipy}
Virtanen, P., Gommers, R., Oliphant, T.~E., {et~al.} 2020, Nature Methods, 17,
  261, \dodoi{10.1038/s41592-019-0686-2}

\bibitem[{{Webbink}(1984)}]{Webbink1984}
{Webbink}, R.~F. 1984, \apj, 277, 355, \dodoi{10.1086/161701}

\bibitem[{Wolf \& Schwab(2017)}]{pymesareader}
Wolf, B., \& Schwab, J. 2017, wmwolf/py\_mesa\_reader: Interact with MESA
  Output, 0.3.0,  Zenodo, \dodoi{10.5281/zenodo.826958}

\bibitem[{{Wu} {et~al.}(2020){Wu}, {Everson}, {Schneider}, {Podsiadlowski}, \&
  {Ramirez-Ruiz}}]{Wu2020}
{Wu}, S., {Everson}, R.~W., {Schneider}, F. R.~N., {Podsiadlowski}, P., \&
  {Ramirez-Ruiz}, E. 2020, \apj, 901, 44, \dodoi{10.3847/1538-4357/abaf48}

\bibitem[{{Yarza} {et~al.}(2022){Yarza}, {Razo Lopez}, {Murguia-Berthier},
  {Everson}, {Antoni}, {MacLeod}, {Soares-Furtado}, {Lee}, \&
  {Ramirez-Ruiz}}]{Yarza2022}
{Yarza}, R., {Razo Lopez}, N., {Murguia-Berthier}, A., {et~al.} 2022, arXiv
  e-prints, arXiv:2203.11227.
\newblock \doarXiv{2203.11227}

\end{thebibliography}
